\documentclass[showpacs,prb,aps,twocolumn,amsmath]{revtex4-1}
\usepackage{feynmp}
\usepackage[dvips]{graphicx,color}
\usepackage{longtable}
\usepackage{amsmath}
\usepackage{bm}

\begin{document}
\title{Impurity Effects on the Nodal Structure of Anisotropic
Superconductors}

\author{Naoyuki Miyata}

\affiliation{Toyama Industrial Technology Center,\\
150, Futagami, Takaoka, Toyama, 933-0981, Japan}

\date{\today}

\begin{abstract}
 For anisotropic superconductors, the gap function belongs to an
 irreducible representation of the point group of crystals,
 which determines the nodal structure of the gap function. Impurity effects on anisotropic superconductors
 have been treated by the Born or T-matrix approximation by which only
 a partial sum of the pertutbation series is calculated. Instead, we
 take into acount all terms of the
 perturbation series. As a result, we find that the introduction of nonmagnetic
 impurities does not change the nodal structure in all cases. We also
 find how to obtain nodes of the gap function and show the list for
 $O_{h}$, $D_{4h}$, and $D_{6h}$.
\end{abstract}
\pacs{74.20.Rp,}
\maketitle

\section{Introduction}
Measurements of direction dependent specific heat
and direction dependent thermal conductivity has enabled us to
investigate concrete nodal structures of anisotropic superconductors. 
For anisotropic superconductors, the gap function is considered to
belong to an irreducible representation of the point group of crystals,
which determines the nodal structure of the gap
function\cite{vol,
uedarice,uedarice2}.  
The nodal structure of the gap function has strong influence on
low temperature behaviors of physical quantities, for example, specific
heat and NMR relaxation rate, which are called the power-laws. Thus, from qualitative point
of view, it is more important to
investigate nodal structures than to culculate concrete
values. Furthurmore, since what an irreducible representation the gap
function belongs to has close connection with pairing interaction, its
identification is crucial for understanding the pairing mechanism.

For conventional s-wave superconductors, low temperature behaviors are insensitive to small
concentrations of nonmagnetic impurities (Anderson's Theorem\cite{anderson}). However,
it is not the case with anisotropic superconductors. Particularly, for
gaps with line nodes, density of states (DOS) at the Fermi energy is shifted
to nonzero value, and therefore low temperature behaviors are modified by
arbitrarily small impurity concentrations\cite{gorkovkalugin,
uedarice2}. Thus, impurity effects
play an important roll to identify the representation of the gap function.

In this paper, we deal with nonmagnetic impurity effects for anisotropic superconductors. We assume that the gap function as an order parameter is
obtained by the generalized Ginzburg-Landau theory (GL).
Previously, 
the correction of self-energy part in the Dyson
equation (Gor'kov equation) due to impurities has been treated by the Born
or T-matrix approximation. Instead, we
take into acount all terms of perturbation series of self-energy part,
not as an approximation. We prove that the introduction of nonmagnetic
impurities does not change the representation of the gap function in
section IV. We also show how to obtain nodes of the gap function  and
show the list for $O_{h}$, $D_{6h}$, and $D_{4h}$ in
appendixes. 
\section{Model}
We consider a mean field approximated Hamiltonian which includes
nonmagnetic impurity term,
\begin{align}
H=&\sum_{\bm{k},s}\epsilon(\bm{k})a_{\bm{k}s}^\dagger a_{\bm{k}s}\nonumber\\
&+\frac{1}{2}\sum_{\bm{k},\bm{k}^\prime,s_i}
(\Delta_{s_1s_2}(\bm{k})a_{\bm{k}s_1}^\dagger
 a_{-\bm{k}s_2}^\dagger
-\Delta_{s_1s_2}^{\ast}(-\bm{k})a_{-\bm{k}s_1}
 a_{\bm{k}s_2})\nonumber \\
&+\frac{1}{V}\sum_{\bm{k},\bm{k}^\prime,s}V(\bm{k}-\bm{k}^\prime)\delta\tilde{\rho}(\bm{k}-\bm{k}^\prime)a_{\bm{k} s}^\dagger
 a_{\bm{k}^\prime s},
 \end{align}
where
\begin{align}
&\Delta_{ss^\prime}(\bm{k})\equiv-\sum_{\bm{k}^\prime,s_1,s_2}V_{s^\prime
 ss_1s_2}(\bm{k},\bm{k}^\prime)\langle a_{\bm{k}^\prime s_1}
 a_{-\bm{k}^\prime s_2}\rangle.\label{eq:delta}
\end{align}
Here, $\bm{k}$ is a wave number of electrons,
$s$ is pseudospin of electrons, $\epsilon(\bm{k})$ is
the band energy, $V_{s_1s_2s_3s_4}(\bm{k},\bm{k}^\prime)$ is
an effective electron-electron interaction,
$V$ is volume
of the system, and $a_{\bm{k}s}$ ($a_{\bm{k}s}^\dagger$) are annihilation (creation) operators. 
$\Delta_{ss^\prime}(\bm{k})$ is the gap function.  The last term is nonmagnetic impurity term\cite{ag1, gorkov}.
By $\langle \cdots \rangle$ we mean
not only expectation value at temperature $T$, but also disorder-averaging.

Then, equations which temperature Green's functions satisfy (Gor'kov
equation) are

\begin{align} \hat{G}(\bm{k},i\omega_{n})
=&\hat{G_{0}}(\bm{k},i\omega_{n})\nonumber \\
 -&\hat{G_{0}}(\bm{k},i\omega_{n})\hat{\Sigma}^{(1)}(\bm{k},i\omega_{n})\hat{G}(\bm{k},i\omega_{n})\\
-&\hat{G_{0}}(\bm{k},i\omega_{n})\hat{\Sigma}^{(2)}(\bm{k},i\omega_{n})\hat{F}(\bm{k},-i\omega_{n})^{\dagger}\nonumber
,
\\
\hat{F}(\bm{k},-i\omega_{n})^{\dagger}
=&-\hat{G_{0}}(-\bm{k},i\omega_{n})^{\ast}\hat{\Sigma}^{(2)}(-\bm{k},i\omega_{n})^{\ast}\hat{G}(\bm{k},i\omega_{n})\nonumber
\\-&\hat{G_{0}}(-\bm{k},i\omega_{n})^{\ast}\hat{\Sigma}^{(1)}(-\bm{k},i\omega_{n})^{\ast}\hat{F}(\bm{k},-i\omega_{n})^{\dagger},
\label{eq:gorkovf}
\end{align}
where $\hat{G}(\bm{k},i\omega_{n})$ and $\hat{F}(\bm{k},i\omega_{n})$
are normal and anomalous temperature Green's functions of matrix form
respectively (subscript ``0'' means that for free fermion system),
$\omega_{n}$ is the Matsubara frequency, and $\hat{\Sigma}^{(1)}(\bm{k},i\omega_{n})$ and
$\hat{\Sigma}^{(2)}(\bm{k},i\omega_{n})$ are the proper self-energy
parts.

 Note that $\hat{F}(\bm{k},-i\omega_{n})^{\dagger}$
is not merely complex conjugate of $\hat{F}(\bm{k},-i\omega_{n})$, but
hermitian conjugate of $\hat{F}(\bm{k},-i\omega_{n})$ which differs from
that in the appendix of ref. 8. Throughout this paper, we take an
appropriate regularization if needed.

\begin{figure}
 \begin{center}
  \includegraphics{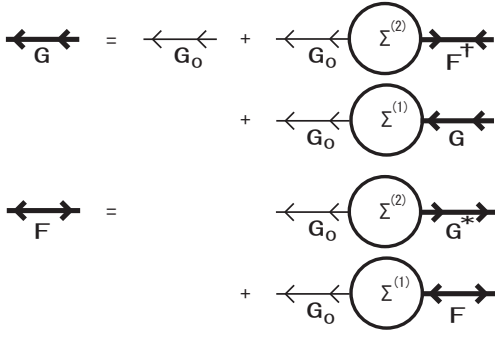}
 \end{center}
\caption{Gor'kov equation}
\label{gorkov}
\end{figure}

\section{Properties of the Gap Function of Clean Superconductors}
If temperature $T$ is very close to a transition point $T_{c}$, the gap
equation becomes a linear equation. Therefore while $T$ is close enough to $T_{c}$, the
gap function is a combination of basis functions of an irreducible
representation $\Gamma^{\alpha}$ of the group $G$ (the largest group which does not change
lattice), such as 
\begin{align}
 \hat{\Delta}(\bm{k})=\sum_{i}c_{i}\hat{\Delta}_{i}^{(\alpha)}(\bm{k}).
\end{align}
$\hat{\Delta}_{i}^{(\alpha)}(\bm{k})$ is i-th basis of irreducible
representation $\Gamma^{\alpha}$. 
These coefficients $\{c_{i}\}$ are determined by GL.
Note that basis functions are the ones for following transformation which
forms faithful representation of $G$:
\begin{align}
 \hat{U}_{R}\hat{\Delta}_{i}^{(\alpha)}(R^{-1}\bm{k})\hat{U}_{R}^{T}
=\sum_{j}\hat{\Delta}_{j}^{(\alpha)}(\bm{k})D_{ji}^{(\alpha)}(R)\label{eq:deltai}\\
\hat{U}_{R}\equiv \hbox{exp}(-i\hat{\bm{\sigma}}\cdot \bm{n} \theta/2)
\end{align}
$R$ is a transformation of $G$. $\hat{\bm{\sigma}}$ is the Pauli's
spin matrix. $\theta$, $\bm{n}$ are the rotation angle and the unit
vector which represent the direction of the rotational part (in the sense of
right handed screw) of the
transformation $R^{-1}$. $D_{ij}^{(\alpha)}(R)$ is an (ij)
component of a matrix of irreducible representation $\Gamma^{\alpha}$ of
$G$ for $R$. Transformations for pseudospins are derived from the fact
that due to spin-orbit coupling, interaction $V_{s_1s_2s_3s_4}(\bm{k},\bm{k}^\prime)$ is not
invariant under rotations of wavenumbers unless we rotate pseudospins
simultaneously. 

We can write the gap function with $\psi(\bm{k})$ or
$\bm{d}(\bm{k})$ corresponding to singlet (even parity) or triplet (odd
parity), such as $\hat{\Delta}(\bm{k})=i\sigma_{y}\psi(\bm{k})$,
$\hat{\Delta}(\bm{k})=i(\bm{d}(\bm{k})\cdot\bm{\sigma})\sigma_{y}$,
respectively, where $\bm{\sigma}$ and $\sigma_{y}$ are the Pauli's spin matrices. 
By comparison with Eq.(\ref{eq:deltai}), we find that, instead of
$\hat{\Delta}_{i}(\bm{k})$, we can treat $\psi(\bm{k})$ or
$\bm{d}(\bm{k})$ which satisfy:
\begin{align}
\hbox{singlet:}\nonumber \\
&\psi(\bm{k})=\sum_{i}c_{i}\psi_{i}^{(\alpha)}(\bm{k}),\\
&\psi_{i}^{(\alpha)}(R^{-1}\bm{k})=\sum_{j}\psi_{j}^{(\alpha)}(\bm{k})D_{ji}^{(\alpha)}(R),\\
\hbox{triplet:}\nonumber \\
&\bm{d}(\bm{k})=\sum_{i}c_{i}\bm{d}_{i}^{(\alpha)}(\bm{k}),\\
&\tilde{R}^{-1}\bm{d}_{i}^{(\alpha)}(R^{-1}\bm{k})=\sum_{j}\bm{d}_{j}^{(\alpha)}(\bm{k})D_{ji}^{(\alpha)}(R),
\end{align} 
where $\tilde{R}$ is a rotation matrix which does not have the parity
transformation part of $R$. Throughout this paper, we take the basis
functions $\psi_{i}^{(\alpha)}(\bm{k})$ and $\bm{d}_{i}^{(\alpha)}(\bm{k})$
to be real because accidental degeneracy may not occur.

Now, the quasiparticle energy depends on $\hat{\Delta}(\bm{k})$ as
\begin{align}
\hbox{singlet:}\nonumber \\
 &E_{\bm{k}}\equiv
\sqrt{\epsilon(\bm{k})^{2}+\frac{1}{2}\hbox{tr}
 \hat{\Delta}(\bm{k})\hat{\Delta}(\bm{k})^{\dagger}}
=\sqrt{\epsilon(\bm{k})^{2}+|\psi({\bm{k}})|^{2}},\\
\hbox{triplet:}\nonumber \\
&E_{\bm{k}\pm}\equiv
\sqrt{\epsilon(\bm{k})^{2}+|\bm{d}(\bm{k})|^{2}\pm|\bm{q}(\bm{k})|},
\end{align}
where $\bm{q}(\bm{k})\equiv
i\bm{d}(\bm{k})\times\bm{d}(\bm{k})^{\ast}$. 
$\hat{\Delta}(\bm{k})$ has zero points derived from its belonging
irreducible representation (See Appendix B). The zero points make nodes of the excitation
energy on the Fermi surface. 

\section{Superconductors in the presence of Nonmagnetic Impurities}
\begin{figure}[htbp]
 \begin{center}
  \includegraphics{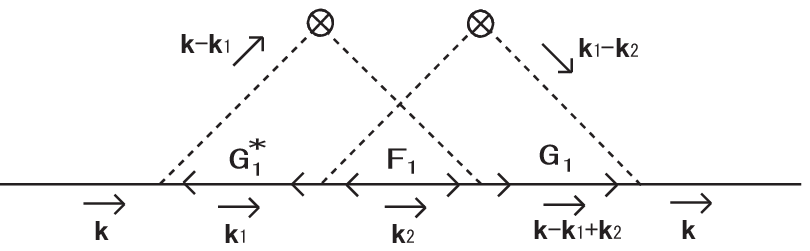}
 \end{center}
\caption{}
\label{gfgstar}
\end{figure}

Since we can find that the quantity in the presence of nonmagnetic impurities corresponding to the
gap function of clean superconductors is the proper self-energy part of anomalous type
$\hat{\Sigma}^{(2)}(\bm{k},i\omega_{n})$, let us consider $\hat{\Sigma}^{(2)}(\bm{k},i\omega_{n})$.
For convenience, we first consider a term
$\hat{G}_{1}\hat{F}_{1}\hat{G}_{1}^{\ast}$ of perturbation
series of $\hat{\Sigma}^{(2)}(\bm{k},i\omega_{n})$ (See Fig.\ref{gfgstar}).
Note that the subscript ``1'' means the exact Green's functions in the
absence of impurities.

The transformation of $\hat{\Sigma}^{(2)}(\bm{k},i\omega_{n})$ such as
$\hat{\Sigma}^{(2)}(\bm{k},\omega)\to
\hat{U}_{R}\hat{\Sigma}^{(2)}(R^{-1}\bm{k},i\omega_{n})\hat{U}_{R}^{T}$ leads
to corresponding transformation of
$\hat{G}_{1}\hat{F}_{1}\hat{G}_{1}^{\ast}$ term included in
$\hat{\Sigma}^{(2)}(\bm{k},i\omega_{n})$ such as 

\begin{align}
&n_{i}^{2}\sum_{\bm{k}_{1},\bm{k}_{2}}V(\bm{k}-\bm{k}_{1})^{2}V(\bm{k}_{1}-\bm{k}_{2})^{2}\nonumber
 \\
&\times \hat{G}_{1}(\bm{k}-\bm{k}_{1}+\bm{k}_{2},i\omega_{n})\hat{F}_{1}(\bm{k}_{2},i\omega_{n})\hat{G}_{1}^{\ast}(-\bm{k}_{1},i\omega_{n})\\
\to
&
 n_{i}^{2}\sum_{\bm{k}_{1},\bm{k}_{2}}V(R^{-1}\bm{k}-\bm{k}_{1})^{2}V(\bm{k}_{1}-\bm{k}_{2})^{2}\nonumber
 \\
&\times\hat{U}_{R}\hat{G}_{1}(R^{-1}\bm{k}-\bm{k}_{1}+\bm{k}_{2},i\omega_{n})\nonumber
 \\
&\times\hat{F}_{1}(\bm{k}_{2},i\omega_{n})\hat{G}_{1}^{\ast}(-\bm{k}_{1},i\omega_{n})\hat{U}_{R}^{T}\\
=&n_{i}^{2}\sum_{R\bm{k}_{1},R\bm{k}_{2}}V(R^{-1}\bm{k}-R^{-1}\bm{k}_{1})^{2}V(R^{-1}\bm{k}_{1}-R^{-1}\bm{k}_{2})^{2} \nonumber
 \\
&\times\hat{U}_{R}\hat{G}_{1}(R^{-1}\bm{k}-R^{-1}\bm{k}_{1}+R^{-1}\bm{k}_{2},i\omega_{n})\nonumber
 \\
&\times\hat{F}_{1}(R^{-1}\bm{k}_{2},i\omega_{n})\hat{G}_{1}^{\ast}(-R^{-1}\bm{k}_{1},i\omega_{n})\hat{U}_{R}^{T}
\\
=&n_{i}^{2}\sum_{\bm{k}_{1},\bm{k}_{2}}V(R^{-1}(\bm{k}-\bm{k}_{1}))^{2}V(R^{-1}(\bm{k}_{1}-\bm{k}_{2}))^{2}\nonumber
\\
&\times\hat{U}_{R}\hat{G}_{1}(R^{-1}(\bm{k}-\bm{k}_{1}+\bm{k}_{2}),i\omega_{n})\hat{U}_{R}^{\dagger}\nonumber
 \\
&\times\hat{U}_{R}\hat{F}_{1}(R^{-1}\bm{k}_{2},i\omega_{n})\hat{U}_{R}^{T}\nonumber
 \\
&\times\hat{U}_{R}^{\ast}\hat{G}_{1}^{\ast}(-R^{-1}\bm{k}_{1},i\omega_{n})\hat{U}_{R}^{T},\\
&\hbox{now, if $V(\bm{k})$ satisfies $V(R^{-1}\bm{k})=V(\bm{k})$, 
}
\\
=&n_{i}^{2}\sum_{\bm{k}_{1},\bm{k}_{2}}V(\bm{k}-\bm{k}_{1})^{2}V(\bm{k}_{1}-\bm{k}_{2})^{2}\nonumber
 \\
&\times\hat{U}_{R}\hat{G}_{1}(R^{-1}(\bm{k}-\bm{k}_{1}+\bm{k}_{2}),i\omega_{n})\hat{U}_{R}^{\dagger}\nonumber
 \\
&\times\hat{U}_{R}\hat{F}_{1}(R^{-1}\bm{k}_{2},i\omega_{n})\hat{U}_{R}^{T}\nonumber
 \\
&\times\hat{U}_{R}^{\ast}\hat{G}_{1}^{\ast}(-R^{-1}\bm{k}_{1},i\omega_{n})\hat{U}_{R}^{T},
\end{align}

where $n_{i}$ is the impurity concentration. This implies that transformation such as $\hat{\Sigma}^{(2)}(\bm{k},i\omega_{n})\to
\hat{U}_{R}\hat{\Sigma}^{(2)}(R^{-1}\bm{k},i\omega_{n})\hat{U}_{R}^{T}$ is
equivalent to the transformation of Green's functions in
$\hat{\Sigma}^{(2)}(\bm{k},i\omega_{n})$ such as

\begin{align}
\hat{G}_{1}(\bm{k},i\omega_{n})\to
 \hat{U}_{R}\hat{G}_{1}(R^{-1}\bm{k},i\omega_{n})\hat{U}_{R}^{\dagger},\label{eq:trans1}
\\
\hat{F}_{1}(\bm{k},i\omega_{n})\to 
\hat{U}_{R}\hat{F}_{1}(R^{-1}\bm{k},i\omega_{n})\hat{U}_{R}^{T},
\\
\hat{G}_{1}^{\ast}(-\bm{k},i\omega_{n})\to
\hat{U}_{R}^{\ast}\hat{G}_{1}^{\ast}(-R^{-1}\bm{k},i\omega_{n})\hat{U}_{R}^{T},
\\
\hat{F}_{1}^{\dagger}(\bm{k},-i\omega_{n})\to 
\hat{U}_{R}^{\ast}\hat{F}_{1}^{\dagger}(R^{-1}\bm{k},-i\omega_{n})\hat{U}_{R}^{\dagger}.\label{eq:trans2}
\end{align}
Although the current example does not have
$\hat{F}_{1}^{\dagger}(\bm{k},-i\omega_{n})$, we can always treat
general terms like this. The reason for this is as follows. 

Combinations of adjoining two Green's functions appearing in
$\hat{\Sigma}^{(2)}(\bm{k},i\omega_{n})$ are restricted as left raw
below because of particle number conservation at each point. And we can
insert spin rotation matrices into two Green's functions as right raw
below. Moreover, it is always possible to transform internal wave numbers
as well as
$\hat{G}_{1}\hat{F}_{1}\hat{G}_{1}^{\ast}$ because internal wave numbers
are merely integral variables . 
\begin{center}

$\hat{G}_{1}\hat{G}_{1}\to \hat{G}_{1}\hat{U}^{\dagger}\hat{U}\hat{G}_{1},$\\

$\hat{G}_{1}\hat{F}_{1}\to \hat{G}_{1}\hat{U}^{\dagger}\hat{U}\hat{F}_{1},$\\

$\hat{F}_{1}\hat{G}_{1}^{\ast}\to \hat{F}_{1}\hat{U}^{T}\hat{U}^{\ast}\hat{G}_{1}^{\ast},$\\

$\hat{F}_{1}\hat{F}_{1}^{\dagger}\to \hat{F}_{1}\hat{U}^{T}\hat{U}^{\ast}\hat{F}_{1}^{\dagger},$\\

$\hat{G}_{1}^{\ast}\hat{G}_{1}^{\ast}\to \hat{G}_{1}^{\ast}\hat{U}^{T}\hat{U}^{\ast}\hat{G}_{1}^{\ast},$\\

$\hat{G}_{1}^{\ast}\hat{F}_{1}^{\dagger}\to \hat{G}_{1}^{\ast}\hat{U}^{T}\hat{U}^{\ast}\hat{F}_{1}^{\dagger},$\\

$\hat{F}_{1}^{\dagger}\hat{G}_{1}\to \hat{F}_{1}^{\dagger}\hat{U}^{\dagger}\hat{U}\hat{G}_{1},$\\

$\hat{F}_{1}^{\dagger}\hat{F}_{1}\to
\hat{F}_{1}^{\dagger}\hat{U}^{\dagger}\hat{U}\hat{F}_{1}.$\\

\end{center}
Therefore examination on what irreducible representations these Green's functions
have for above transformations (\ref{eq:trans1})-(\ref{eq:trans2}) will make it clear that what irreducible representations
$\hat{\Sigma}^{(2)}(\bm{k},i\omega_{n})$ has for above transformations for $\hat{\Sigma}^{(2)}(\bm{k},i\omega_{n})$.
\\

Here, we define $G^{\prime}$, $\Gamma^{\prime}_{\beta}$, and
$D^{\prime}_{\beta}$. By $G^{\prime}$ we mean the maximal subgroup of $G$ where
the quasiparticle energy of clean superconductor is invariant. We can find
that $G^{\prime}$ is
not only maximal but also the largest. Because if $G^{\prime\prime}$
where the excitation energy is invariant is not contained by
$G^{\prime}$, the quasiparticle energy is also invariant under the subgroup
which is generated from $G^{\prime}\cup G^{\prime\prime}(\supset G^{\prime})$, this
contradicts the fact that $G^{\prime}$ is maximal. $\Gamma^{\prime}_{\beta}$ is an irreducible
representation subspace of $G^{\prime}$.  

$D^{\prime}_{\beta}$ is a subspace
of $\Gamma^{\prime}_{\beta}$ generated from elements which have a form of
 $\sum_{i}c_{i}^{\prime}\psi_{i}^{\prime(\beta)}=\sum_{i}c_{i}\psi_{i}^{(\alpha)}$.
The coefficients $\{c_{i}\}$ are determined by GL. $\{c_{i}^{\prime}\}$ are determined uniquely from $\{c_{i}\}$.

Let us examine what irreducible representations Green's functions have
under above assumption. In what follows, $\hat{\Delta}(\bm{k})$ is called unitary if the
product $\hat{\Delta}(\bm{k})\hat{\Delta}(\bm{k})^{\dagger}$
is proportional to the unit matrix, otherwise it is called
nonunitary. Note that only triplet case can be nonunitary.

It can be proved that regardless of the absence or the presence of
 impurities, $\hat{\Delta}(\bm{k})\in
 D^{\prime}_{\beta}\subset \Gamma^{\prime}_{\beta}$ until level crossing
 (including by another bifurcation from the solution) occurs for any GL
 solution.\cite{miyata}. In what follows, we shall consider two cases (i) and (ii) separately.

(i) 
In case order parameter coefficients $\{c_i\}$ obtained as a solution of GL are all real (complex number multiplied overall
is neglected), only unitary case applies to this case.

Exact Green's functions of clean superconductors are 
\begin{align}
&\hat{G}_{1}(\bm{k},i\omega_{n})
=-\frac{i\omega_{n}+\epsilon(\bm{k})}{\omega_{n}^{2}
+E_{\bm{k}}^{2}}\hat{\sigma_{0}},
\\
&\hat{F}_{1}(\bm{k},i\omega_{n})
=-\frac{\hat{\Delta}(\bm{k})}{\omega_{n}^{2}+E_{\bm{k}}^{2}}.
\end{align}
Because the excitation energy of clean superconductors and the
band
energy satisfies $E_{\bm{k}^\prime}=E_{\bm{k}}$,
$\epsilon(\bm{k}^\prime)=\epsilon(\bm{k})$

 under
transformations of $G^{\prime}$ respectively, Green's functions
are transformed by $G^{\prime}$ as
\begin{align}
 \hat{U}_{R}\hat{G}_{1}(R^{-1}\bm{k},i\omega_{n})\hat{U}_{R}^\dagger
&=-\frac{i\omega_{n}+\epsilon(R^{-1}\bm{k})}{\omega_{n}^{2}+E_{R^{-1}\bm{k}}^{2}}\hat{U}_{R}\hat{\sigma_{0}}\hat{U}_{R}^\dagger\notag
 \\
&=\hat{G}_{1}(\bm{k},i\omega_{n}),
\\
\hat{U}_{R}\hat{F}_{1}(R^{-1}\bm{k},i\omega_{n})\hat{U}_{R}^T
&=\frac{\hat{U}_{R}\hat{\Delta}(R^{-1}\bm{k})\hat{U}_{R}^T}{\omega_{n}^{2}+E_{R^{-1}\bm{k}}^{2}}\notag\\
&=\frac{\hat{U}_{R}\hat{\Delta}(R^{-1}\bm{k})\hat{U}_{R}^T}{\omega_{n}^{2}+E_{\bm{k}}^{2}}.
\end{align}
Hence, 
\begin{align}
\hat{G}_{1}(\bm{k},i\omega_{n})\in\Gamma^{\prime+}_{1},\\
\hat{F}_{1}(\bm{k},i\omega_{n})\in D^{\prime}_{\beta}.
\end{align}
Because $\{c_i\}$ are all real,
\begin{align}
[D^{\prime}_{\beta}\otimes
D^{\prime\ast}_{\beta}]_{AS}=0, \label{eq:kantankatei}
\end{align}
where $[\cdots]_{AS}$ means antisymmetric tensor product. 
Furthermore, since we can find that the condition which the excitation energy
for unitary case is
invariant under $G^{\prime}$ is equivalent to 
\begin{align}
 [D^{\prime}_{\beta}\otimes D^{\prime\ast}_{\beta}]_{S}\subset \Gamma^{\prime+}_{1}.\label{eq:energy} 
\end{align}

Therefore,
\begin{align}
D^{\prime}_{\beta}\otimes
D^{\prime\ast}_{\beta}
&=[D^{\prime}_{\beta}\otimes
D^{\prime\ast}_{\beta}]_{S}+[D^{\prime}_{\beta}\otimes
D^{\prime\ast}_{\beta}]_{AS}\nonumber \\
&\subset \Gamma^{\prime+}_{1}+0=\Gamma^{\prime+}_{1}. \label{eq:kantan}
\end{align}
Now, the representation of each term of $\hat{\Sigma}^{(2)}(\bm{k},i\omega_{n})$ is
\begin{align}
\underbrace{D^{\prime}_{\beta}\otimes
D^{\prime\ast}_{\beta}}_{\Gamma^{\prime+}_{1}}
\otimes
\underbrace{D^{\prime}_{\beta}\otimes
D^{\prime\ast}_{\beta}}_{\Gamma^{\prime+}_{1}}
\otimes\cdots\otimes
\underbrace{D^{\prime}_{\beta}\otimes
D^{\prime\ast}_{\beta}}_{\Gamma^{\prime+}_{1}}
\otimes
D^{\prime}_{\beta}
\subset  D^{\prime}_{\beta}.
\end{align}
Here, we neglect $\hat{G}_{1}(\bm{k},i\omega_{n})$ and
$\hat{G}_{1}^{\ast}(\bm{k},i\omega_{n})$ since they belong to $\Gamma^{\prime+}_{1}$. 

And the facts that $\hat{F}(\bm{k},i\omega_{n})$ and $\hat{F}^{\dagger}(\bm{k},-i\omega_{n})$ appear in
turn, and the number of $\hat{F}(\bm{k},i\omega_{n})$ is always greater than $\hat{F}^{\dagger}(\bm{k},-i\omega_{n})$ by one are used.

Because all terms appearing in $\hat{\Sigma}^{(2)}(\bm{k},i\omega_{n})$ belong
to the same $D^{\prime}_{\beta}$ as the original 
$\hat{\Delta}(\bm{k})$, if equation (\ref{eq:kantan})
holds, $\hat{\Sigma}^{(2)}(\bm{k},i\omega_{n})$ belongs to the same
$D^{\prime}_{\beta}$ as the original $\hat{\Delta}(\bm{k})$ under $G^{\prime}$. It should be noted
that in the present case, in the
table in Sigrist-Ueda\cite{sigristueda},
 $\Gamma^{\prime}_{\beta}$ are all 1D irreducible representation, hence
$D^{\prime}_{\beta}=\Gamma^{\prime}_{\beta}$.
\\

(ii) 
Whereas, in the case of complex coefficients $\{c_i\}$,
equation (\ref{eq:kantan}) does not hold. Because there is no such 1D representation
of $G$, it is enough for us to prove the case of 2D and 3D.

For $G^{\prime}$, any gap function which has complex coefficients is 2D irreducible
representation\cite{miyata}. Now, 
$[D^{\prime}_{\beta}\otimes D^{\prime\ast}_{\beta}]_{AS}$ is a certain 1D
representation other than $\Gamma^{\prime}_{1+}$ 
because of its dimension and it can be proved that it is not $\Gamma^{\prime+}_{1}$. Hence, we find
\begin{align}
[D^{\prime}_{\beta}\otimes D^{\prime\ast}_{\beta}]_{AS}\subset \Gamma^{\prime}_{1D}, 
\end{align}
where $\Gamma^{\prime}_{1D}$ is defined as a certain 1D representation
other than $\Gamma^{\prime+}_{1}$ which has
$[D^{\prime}_{\beta}\otimes D^{\prime\ast}_{\beta}]_{AS}$ as a subset.

It is worth mentioning that in this case, nonunitary cases arise, and 
it is easy to verify that

\begin{align}
&\hat{G}_{1}(\bm{k},i\omega_{n})\in\Gamma^{\prime+}_{1}+[D^{\prime}_{\beta}\otimes D^{\prime\ast}_{\beta}]_{AS},\\
&\hat{F}_{1}(\bm{k},i\omega_{n})\in D^{\prime}_{\beta},
\end{align}

in nonunitary cases.

In any case of (ii), i.e., unitary or nonunitary,  general term of $\Sigma^{(2)}(\bm{k},i\omega_{n})$
has the following representation:

\begin{align}
 &D^{\prime}_{\beta}+\Gamma^{\prime}_{1D}\otimes D^{\prime}_{\beta}\\
\end{align}
We can find that in all cases, this is $D^{\prime}_{\beta}$, and the solution of the
gap equation is not affected by the introduction of nonmagnetic impurities\cite{miyata}. 
\\

After all, in any case, we can find that
$\Sigma^{(2)}(\bm{k},i\omega_{n})\in D^{\prime}_{\beta}$.

\section{Summary and Discussion}

We proved that the introduction of nonmagnetic impurities does not
change the nodal structure of anisotropic superconductors. We also
show why zeros of $\Sigma^{(2)}(\bm{k},0)$ are related to the nodal
structure in Appendix A, how to obtain nodal structures in Appendix B
and show them  in the case of $O_{h}$, $D_{6h}$, and $D_{4h}$ in the
table II, III, and IV of Appendix C respectively. Calculations in
Appendix C and D were done by using MATHEMATICA.

Although theoretical culculations for anisotropic s-wave gaps
show that the gap anisotropy is smeared out by introducing nonmagnetic impurities\cite{borkowski}, and
anisotropic s-wave superconductivity is therefore anticipated to explain
experimental evidences of YNi$_{2}$B$_{2}$C\cite{qingshan,sgnode,maki}, 
it is no wonder that
the nodes disappear because the nodes are not
the ones as an irreducible representation.

We should pay attention to that even if nodes of gaps disappear due to
impurities, we do not say anything about the amplitudes of gaps. Hence, 
the fact obtained in this paper does not contradict  
the fact that the axial-like gap (gap with point node(s)) and the polar-like gap (gap with line node(s)) are suppressed partially by
introducing nonmagnetic impurities in the case of the Born approximation\cite{uedarice2,gorkovkalugin}, and in the case
of the T-matrix
approximation\cite{smv,hvw,haas}. Especially, a theoretical culculation by
Haas $et$ $al$.\cite{haas} and an experiment by Hashimoto $et$
$al$.\cite{hashimoto} agrees with our theory, although these are too
simple to be evidences.


\section*{Acknowledgement}
This work is based on the prior research by Takahiro Aoyama, Masatoshi Sato,
and Mahito Kohmoto. The author acknowledges them for helpful discussions.
\appendix

\section{The Existence of the nodes}

From Gor'kov equation, we can derive that

\begin{align}
&((i\omega_{n}-\epsilon(\bm{k}))\hat{\sigma}_{0}-\hat{\Sigma}^{(1)}(\bm{k},i\omega_{n})+\hat{\Sigma}^{(2)}(\bm{k},i\omega_{n})\notag
 \\
&\times((-i\omega_{n}-\epsilon(\bm{k}))\hat{\sigma_{0}}+\hat{\Sigma}^{(1)}(-\bm{k},i\omega_{n})^{\ast})^{-1}\notag
 \\
&\times\hat{\Sigma}^{(2)}(-\bm{k},i\omega_{n})^{\ast})\hat{G}(\bm{k},i\omega_{n})\notag\\
&=\hat{\sigma}_{0}
\end{align}

Putting $\omega$ to $0$, we can obtain zero energy excitation. Consider
a point where $\Sigma^{(2)}(\bm{k},0)=0$. At that point, if
$((-\epsilon(\bm{k}))\hat{\sigma_{0}}+\hat{\Sigma}^{(1)}(-\bm{k},0)^{\ast})^{-1}$
is divergent more strongly than two $\Sigma^{(2)}(\bm{k},0)$s on both
side, the Green's function has no pole there. In this case, there is no
node.

If it is divergent less
strongly than or equal to two $\Sigma^{(2)}(\bm{k},0)$s on both
side, or not at all divergent, the equation becomes

\begin{align}
&((\epsilon(\bm{k})\hat{\sigma}_{0}+\hat{\Sigma}^{(1)}(\bm{k},0))\hat{G}(\bm{k},0)\notag\\
&=-\hat{\sigma}_{0},
\end{align}

where we renormalize the residue into $\Sigma^{(1)}(\bm{k},0)$ if it
exists.

Moreover, $\Sigma^{(1)}(\bm{k},0)^{\dagger}=\Sigma^{(1)}(\bm{k},0)$
because the transformation
$\Sigma^{(1)}(\bm{k},0)\to\Sigma^{(1)}(\bm{k},0)^{\dagger}$ causes
transformations on internal Green's functions such as,
\\

$(-\hat{G}_{1}(\bm{k}_{1},0))(-\hat{G}_{1}(\bm{k}_{2},0))\to $

$(-\hat{G}_{1}(\bm{k}_{2},0))^{\dagger}(-\hat{G}_{1}(\bm{k}_{1},0))^{\dagger}$

$=(-\hat{G}_{1}(\bm{k}_{2},0))(-\hat{G}_{1}(\bm{k}_{1},0)),$
\\

$(-\hat{G}_{1}(\bm{k}_{1},0))\hat{F}_{1}(\bm{k}_{2},0)\to$

$ \hat{F}_{1}(\bm{k}_{2},0)^{\dagger}(-\hat{G}_{1}(\bm{k}_{1},0))^{\dagger}$

$=\hat{F}_{1}(\bm{k}_{2},0)(-\hat{G}_{1}(-\bm{k}_{1},0)),$\\

$\hat{F}_{1}(\bm{k}_{1},0)\hat{G}_{1}(-\bm{k}_{2},0)^{\ast}\to$

$ \hat{G}_{1}(-\bm{k}_{2},0)^{\ast}\hat{F}_{1}(\bm{k}_{1},0)^{\dagger}$

$=\hat{G}_{1}(-\bm{k}_{2},0)^{\ast}\hat{F}_{1}(\bm{k}_{1},0)^{\dagger},$\\

$\hat{F}_{1}(\bm{k}_{1},0)\hat{F}_{1}(\bm{k}_{2},0)^{\dagger}\to$

$ \hat{F}_{1}(\bm{k}_{2},0)\hat{F}_{1}(\bm{k}_{1},0)^{\dagger},$\\

$\hat{G}_{1}(-\bm{k}_{1},0)^{\ast}\hat{G}_{1}(-\bm{k}_{2},0)^{\ast}\to$

$\hat{G}_{1}(-\bm{k}_{2},0)^{T}\hat{G}_{1}(-\bm{k}_{1},0)^{T}$

$=\hat{G}_{1}(-\bm{k}_{2},0))^{\ast}\hat{G}_{1}(\bm{k}_{2},0)^{\ast},$\\

$\hat{G}_{1}(-\bm{k}_{1},0)^{\ast}\hat{F}_{1}(\bm{k}_{2},0)^{\dagger}\to$

$ \hat{F}_{1}(\bm{k}_{2},0)\hat{G}_{1}(-\bm{k}_{1},0)^{T}$

$=\hat{F}_{1}(\bm{k}_{2},0)\hat{G}_{1}(-\bm{k}_{1},0)^{\ast},$\\

$\hat{F}_{1}(\bm{k}_{1},0)^{\dagger}(-\hat{G}_{1}(\bm{k}_{2},0))\to$

$ (-\hat{G}_{1}(\bm{k}_{2},0))\hat{F}_{1}(\bm{k}_{1},0)=$

$(-\hat{G}_{1}(\bm{k}_{2},0))\hat{F}_{1}(\bm{k}_{1},0),$\\

$\hat{F}_{1}(\bm{k}_{1},0)^{\dagger}\hat{F}_{1}(\bm{k}_{2},0)\to$

$\hat{F}_{1}(\bm{k}_{2},0)^{\dagger}\hat{F}_{1}(\bm{k}_{1},0).$\\

Note that in $\Sigma^{(1)}(\bm{k},0)$, the number of $F_{1}$ and
$F_{1}^{\dagger}$ is equall. Similar to the argument of
$\Sigma^{(2)}(\bm{k},0)$, the transformation causes a change with dually
corresponding term with no sign change.

Thus, $\Sigma^{(1)}(\bm{k},0)^{\dagger}=\Sigma^{(1)}(\bm{k},0)$, and
$(\epsilon(\bm{k})\hat{\sigma}_{0}+\hat{\Sigma}^{(1)}(\bm{k},0))$ can be
diagonalized by a unitaty matrix.

Let us write the eigenvalues as
$\epsilon(\bm{k})+\Sigma^{(1)}_{i}(\bm{k},0)$, $(i=1,2)$.

$\epsilon(\bm{k})$ changes its sign on the Fermi surface, while
$\Sigma^{(1)}_{i}(\bm{k},0)$ should be slowly varying in the vicinity of the
Fermi surface if impurity concentration is sufficiently
small although this needs to be proved. Therefore, there should be a point near the Fermi surface where
$\epsilon(\bm{k})+\Sigma^{(1)}_{i}(\bm{k},0)=0$. At
that point, zero energy excitation exists.

\section{How to obtain nodes of the gap functions}
The fact that $\Sigma^{(2)}(\bm{k},0)$ has the same parity as the gap
function is almost trivial. This condition and the condition
$\Sigma^{(2)}(-\bm{k},0)^{T}=-\Sigma^{(2)}(\bm{k},0)$ leads to the fact
that $\Sigma^{(2)}(\bm{k},0)$ can be written in the same form as the gap
function, i.e., $\Sigma^{(2)}(\bm{k},0)=i\sigma_{y}\psi(\bm{k})$ or
$\Sigma^{(2)}(\bm{k},0)=i(\bm{d}(\bm{k})\cdot\bm{\sigma})\sigma_{y}$
with certain functions $\psi(\bm{k})$ or $\bm{d}(\bm{k})$,
corresponding to the parity, respectively.

Searching for zeros of $\Sigma^{(2)}(\bm{k},0)$ is
reduced to searching for those of $\psi({\bm{k}})$ or
$\bm{d}(\bm{k})$. The fact
$\Sigma^{(2)}(-\bm{k},0)^{T}=-\Sigma^{(2)}(\bm{k},0)$ is proved in the
latter half of this appendix.\\ 
 
Although there are many sets of basis functions of irreducible
representations, it may occur that any basis functions become zero at
certain points. Here, we consider such zero points.
\\

singlet case:

We write basis functions
$\{\psi_{i}^{(\alpha)}\}_{i=1,2,\cdots,d_{\alpha}}$ of the irreducible representation
$\Gamma^{\alpha}$ ($d_{\alpha}$ dimension) of the point group $G$
in the form:
\begin{align}
\bm{\psi}(\bm{k})=
\begin{pmatrix}
\psi_{1}^{(\alpha)}\\
\psi_{2}^{(\alpha)}\\
\vdots\\
\psi_{d_{\alpha}}^{(\alpha)}
\end{pmatrix}.
\end{align}
And $\bm{\psi}(\bm{k})$ satisfies
\begin{align}
\bm{\psi}(R^{-1}\bm{k})=\hat{D}(R)^{T}\bm{\psi}(\bm{k}),\label{eq:kiyaku}
\end{align}
where $(\hat{D}(R))_{ij}=D_{ij}^{(\alpha)}(R)$ is a representation
matrix of the group element $R$ of the irreducible representation $\Gamma^{\alpha}$.

The equation (\ref{eq:kiyaku}) gives relationships between values of
$\bm{\psi}(\bm{k})$ at
points which is moved by elements of $G$. 

Now, let us consider a point
$\bm{k}$ which is not invariant under any element $R$ of $G$. Then,
$\{R^{-1}\bm{k}\}_{R\in G}$ consists of points of the same number $g$ as
the order of $G$, and values of $\bm{\psi}(\bm{k})$ at $\{R^{-1}\bm{k}\}_{R\in G}$
have relationships. Since there are at most $d_{\alpha}(g-1)$
independent linear equations for $d_{\alpha}g$
variables $\{\bm{\psi}(R^{-1}\bm{k})\}_{R\in G}$as
\begin{align}
\begin{pmatrix}
 \hat{1}& & & & &-\hat{D}(R_{2})^{T}\\
 &\hat{1}& & & &-\hat{D}(R_{3})^{T}\\
 & &\ddots& & &\vdots\\
& & &\hat{1}& & -\hat{D}(R_{g})^{T}
\end{pmatrix}
\begin{pmatrix}
\psi({R_{2}^{-1}\bm{k}})\\
\psi({R_{3}^{-1}\bm{k}})\\
\vdots\\
\psi({R_{g}^{-1}\bm{k}})\\
\psi({\bm{k}})
\end{pmatrix}
=\bm{0},
\end{align}
the solutions are
\begin{align}
\begin{pmatrix}
\psi({R_{2}^{-1}\bm{k}})\\
\psi({R_{3}^{-1}\bm{k}})\\
\vdots\\
\psi({R_{g}^{-1}\bm{k}})\\
\psi({\bm{k}})
\end{pmatrix}
=
\begin{pmatrix}
\begin{pmatrix}
 \hat{D}(R_{2})^{T}\\
 \hat{D}(R_{3})^{T}\\
\vdots\\
\hat{D}(R_{g})^{T}\\
\hat{1}
\end{pmatrix}
\bm{a}
\end{pmatrix}
\end{align}
where $\bm{a}$ is an arbitrary $d_{\alpha}$ dimensional vector, and
$\hat{D}(R_{1})=\hat{1}$ ($d_{\alpha}$ dimensional unit matrix).
Therefore no restriction that forces
$\{\bm{\psi}(R^{-1}\bm{k})\}_{R\in G}$ to be zeros appears.

Next we consider a point $\bm{k}$ which is invariant under some elements of $G$. These points can exist only when they are on rotational axes
or mirror reflection planes. In contrast to previous case, since
$\bm{\psi}(\bm{k})$ with the condition $R^{-1}\bm{k}=\bm{k}$ satisfies
\begin{align}
\bm{\psi}(\bm{k})=\hat{D}(R)^{T}\bm{\psi}(\bm{k})\\
\to(\hat{D}(R)^{T}-\hat{1})\bm{\psi}(\bm{k})=\bm{0},\label{eq:node}
\end{align}
the value of $\bm{\psi}(\bm{k})$
is restricted. 

(i) In the case that there exist $R$ such that the rank of the matrix $(\hat{D}(R)^{T}-\hat{1})$
equals $d_{\alpha}$, that is, determinant of the matrix is not equal to
zero, then $\bm{\psi}(\bm{k})=\bm{0}$. In this case, for any set of
coefficients $\{c_{i}^{\alpha}\}$,
$\sum_{i}c_{i}^{\alpha}\psi_{i}^{\alpha}(\bm{k})=\bm{c}\cdot
\bm{\psi}(\bm{k})=\bm{0}$. For example, $(0,0,1)$ (line) and $[0,0,1]$
(point) of $\Gamma^{5+}$ of
$D_{6h}$ applies to this case. 

(ii) In the case that there is no $R$ such that the rank of the matrix $(\hat{D}(R)^{T}-\hat{1})$
equals $d_{\alpha}$. In this case, if the rank of the matrix for an $R$
equals  $n<d_{\alpha}$, the solutions of
the equation (\ref{eq:node}) must lie in $d_{\alpha}-n$ dimensional
subspace. If $\bm{k}$ has several $R$ which satisies
$R^{-1}\bm{k}=\bm{k}$, the subspace of solutions of the equation is the common
subspace of subspaces for each $R$. $\bm{\psi}(\bm{k})$ is not
always $\bm{0}$ in this case. However, if coefficients $
\bm{c}\equiv
(c_{1}^{\alpha}
c_{2}^{\alpha}
\dots
c_{d_{\alpha}}^{\alpha})^{T}
$
belongs to the orthocomplement of the common subspace, then $\bm{c}\cdot\bm{\psi}(\bm{k})=\sum_{i}c_{i}^{\alpha}\psi_{i}^{\alpha}(\bm{k})=0$.
This means that the gap function is zero at the point $\bm{k}$.
\\
  
triplet case:

The same manner as singlet applies to triplet case except for little complexity.
Basis functions $\{\bm{d}_{i}^{\alpha}\}_{i=1,2,\cdots,d_{\alpha}}$ of
the irreducible representation $\Gamma^{\alpha}$ satisfies
\begin{align}
 \tilde{R}^{-1}\bm{d}_{i}^{(\alpha)}(R^{-1}\bm{k})=\sum_{j}\bm{d}_{j}^{(\alpha)}(\bm{k})D_{ji}^{(\alpha)}(R),
\end{align}
where $\tilde{R}$ is that which parity transformation is reduced from
$R$.
We write
\begin{align}
\bm{d}\equiv
 \begin{pmatrix}
\ \bm{d}_{1}^{(\alpha)}\\
\ \bm{d}_{2}^{(\alpha)}\\
\vdots\\
\ \bm{d}_{d_{\alpha}}^{(\alpha)}
\end{pmatrix}.
\end{align}
In this notation, at the point $\bm{k}$ with the condition
$R^{-1}\bm{k}=\bm{k}$, we have
\begin{align}
\begin{pmatrix}
 \tilde{R}^{-1}& & \\
 &\tilde{R}^{-1}& \\
 & &\ddots \\
\end{pmatrix}
\bm{d}(\bm{k})=
\begin{pmatrix}
 \hat{D}_{11}(R)&\hat{D}_{21}(R)&\cdots\\
 \hat{D}_{12}(R)&\hat{D}_{22}(R)& \\
\vdots& &\ddots\\
\end{pmatrix}
\bm{d}(\bm{k})\\
\to \left\{
\begin{pmatrix}
  \tilde{R}& & \\
 &\tilde{R}& \\
 & &\ddots \\
\end{pmatrix}
\begin{pmatrix}
  \hat{D}_{11}(R)&\hat{D}_{21}(R)&\cdots\\
 \hat{D}_{12}(R)&\hat{D}_{22}(R)& \\
\vdots& &\ddots\\
\end{pmatrix}
-\hat{1}
  \right\}\bm{d}(\bm{k})=\bm{0}, \label{triplet}
\end{align}
where $\hat{D}_{ij}(R)=D_{ij}^{(\alpha)}(R)
\begin{pmatrix}
 1&0&0\\
0&1&0\\
0&0&1
\end{pmatrix}
$
and $\hat{1}$ is a $3d_{\alpha}$ dimensional unit marix. In the case corresponding to (ii) of singlet, we must note that
\begin{align}
 \begin{pmatrix}
  c_{1}^{\alpha}&0&0&c_{2}^{\alpha}&0&0\cdots
 &c_{d_{\alpha}}^{\alpha}&0&0\\
0&c_{1}^{\alpha}&0&0&c_{2}^{\alpha}&0\cdots
 &0&c_{d_{\alpha}}^{\alpha}&0\\
0&0&c_{1}^{\alpha}&0&0&c_{2}^{\alpha}\cdots
 &0&0&c_{d_{\alpha}}^{\alpha}\\
\end{pmatrix}
\begin{pmatrix}
\ \bm{d}_{1}^{(\alpha)}\\
\ \bm{d}_{2}^{(\alpha)}\\
\vdots\\
\ \bm{d}_{d_{\alpha}}^{(\alpha)}
\end{pmatrix}\nonumber \\
=\sum_{i}c_{i}^{\alpha}\bm{d}_{i}^{(\alpha)}(\bm{k}).
\end{align}
Since discussion on triplet case will be the same as singlet case
except
for this, we
abbreviate furthur detail.
\\

The proof for $\Sigma^{(2)}(-\bm{k},0)^{T}=-\Sigma^{(2)}(\bm{k},0)$:\\

A transformation $\Sigma^{(2)}(\bm{k},0)\to
\Sigma^{(2)}(-\bm{k},0)^{T}$ induces transformations on internal Green's functions such as:
\\

$(-\hat{G}_{1}(\bm{k}_{1},0))(-\hat{G}_{1}(\bm{k}_{2},0))\to $

$(-\hat{G}_{1}(-\bm{k}_{2},0))^{T}(-\hat{G}_{1}(-\bm{k}_{1},0))^{T}$

$=\hat{G}_{1}(-\bm{k}_{2},0)^{\ast}\hat{G}_{1}(-\bm{k}_{1},0)^{\ast},$
\\

$(-\hat{G}_{1}(\bm{k}_{1},0))\hat{F}_{1}(\bm{k}_{2},0)\to$

$ \hat{F}_{1}(-\bm{k}_{2},0)^{T}(-\hat{G}_{1}(-\bm{k}_{1},0))^{T}$

$=-\hat{F}_{1}(\bm{k}_{2},0)\hat{G}_{1}(-\bm{k}_{1},0)^{\ast},$\\

$\hat{F}_{1}(\bm{k}_{1},0)\hat{G}_{1}(-\bm{k}_{2},0)^{\ast}\to$

$ \hat{G}_{1}(\bm{k}_{2},0)^{\dagger}\hat{F}_{1}(-\bm{k}_{1},0)^{T}$

$=\hat{G}_{1}(-\bm{k}_{2},0)^{\ast}(-1)\hat{F}_{1}(\bm{k}_{1},0),$\\

$\hat{F}_{1}(\bm{k}_{1},0)\hat{F}_{1}(\bm{k}_{2},0)^{\dagger}\to$

$ \hat{F}_{1}(-\bm{k}_{2},0)^{\ast}\hat{F}_{1}(-\bm{k}_{1},0)^{T}$

$=\hat{F}_{1}(\bm{k}_{2},0)^{\dagger}\hat{F}_{1}(\bm{k}_{1},0),$\\

$\hat{G}_{1}(-\bm{k}_{1},0)^{\ast}\hat{G}_{1}(-\bm{k}_{2},0)^{\ast}\to$

$\hat{G}_{1}(\bm{k}_{2},0)^{\dagger}\hat{G}_{1}(\bm{k}_{1},0)^{\dagger}$

$=(-\hat{G}_{1}(\bm{k}_{2},0))(-\hat{G}_{1}(\bm{k}_{2},0)),$\\

$\hat{G}_{1}(-\bm{k}_{1},0)^{\ast}\hat{F}_{1}(\bm{k}_{2},0)^{\dagger}\to$

$ \hat{F}_{1}(-\bm{k}_{2},0)^{\ast}\hat{G}_{1}(\bm{k}_{1},0)^{\dagger}$

$=(-1)\hat{F}_{1}(-\bm{k}_{2},0)^{\dagger}\hat{G}_{1}(\bm{k}_{1},0),$\\

$\hat{F}_{1}(\bm{k}_{1},0)^{\dagger}(-\hat{G}_{1}(\bm{k}_{2},0))\to$

$ (-\hat{G}_{1}(-\bm{k}_{2},0))^{T}\hat{F}_{1}(-\bm{k}_{1},0)^{\ast}=$

$(-\hat{G}_{1}(\bm{k}_{2},0))(-1)\hat{F}_{1}(\bm{k}_{1},0)^{\dagger},$\\

$\hat{F}_{1}(\bm{k}_{1},0)^{\dagger}\hat{F}_{1}(\bm{k}_{2},0)\to$

$\hat{F}_{1}(-\bm{k}_{2},0)^{T}\hat{F}_{1}(-\bm{k}_{1},0)^{\ast}=$

$\hat{F}_{1}(\bm{k}_{2},0)\hat{F}_{1}(\bm{k}_{1},0)^{\dagger}.$\\

Here, the transformation of external wave number $\bm{k}\to -\bm{k}$ is 
equivalent to the transformation of internal wave number
$\bm{k}_{i}\to-\bm{k}_{i}$.

Thus, by the transformation $\Sigma^{(2)}(\bm{k},0)\to
\Sigma^{(2)}(-\bm{k},0)^{T}$, each term in $\Sigma^{(2)}(\bm{k},0)$ is
multiplied by a factor $-1$, because $-1$ is multiplied ``the summation of the number of
$\hat{F}_{1}$ and $\hat{F}_{1}^{\dagger}$ '' times, and they are always
odd.

In the case of a diagrammatically symmetric term, the transformation
causes only the change of internal wave number indices, and thus $-1$ is multiplied.

In the case of a diagrammatically asymmetric term, the transformation
causes a change with dually corresponding term, and thus $-1$ is
multiplied.

Thus, $\Sigma^{(2)}(-\bm{k},0)^{T}=-\Sigma^{(2)}(\bm{k},0)$.

\section{Table of Nodal Structures of $O_{h}$,
$D_{6h}$, and $D_{4h}$ Anisotropic Superconductors}

We choose basis functions of irreducible representations for
$O_{h}$,$D_{6h}$,and $D_{4h}$ as follows.
\squeezetable
 \begin{longtable}{cl}
\caption{basis functions of irreducible representations for
  $O_{h}$, $D_{6h}$, and $D_{4h}$}\\
\hline \hline 
\multicolumn{1}{c}{$\Gamma$}&
\multicolumn{1}{c}{$\psi(\bm{k}) / \bm{d}(\bm{k})$}
\\ \hline \endhead \hline \endfoot
 \multicolumn{2}{c}{$O_{h}$}
\\
\multicolumn{2}{c}{parity even (singlet)}
\\ 
 $\Gamma^{+}_{1}$ &$\psi^{(1+)}(\bm{k})=1$
\\ 
$\Gamma^{+}_{2}$ &$\psi^{(2+)}(\bm{k})=(k_{x}^{2}-k_{y}^{2})(k_{y}^{2}-k_{z}^{2})(k_{z}^{2}-k_{x}^{2})$
\\ 
$\Gamma^{+}_{3}$ &$\psi^{(3+)}_{1}(\bm{k})=2k_{z}^{2}-k_{x}^{2}-k_{y}^{2}$
\\ 
 &$\psi^{(3+)}_{2}(\bm{k})= \sqrt{3}(k_{x}^{2}-k_{y}^{2})$
\\ 
$\Gamma^{+}_{4}$ &$\psi^{(4+)}_{1}(\bm{k})=k_{y}k_{z}(k_{y}^{2}-k_{z}^{2})$
\\ 
 &$\psi^{(4+)}_{2}(\bm{k})=k_{z}k_{x}(k_{z}^{2}-k_{x}^{2})$
\\ 
 &$\psi^{(4+)}_{3}(\bm{k})=k_{x}k_{y}(k_{x}^{2}-k_{y}^{2})$
\\ 
$\Gamma^{+}_{5}$ &$\psi^{(5+)}_{1}(\bm{k})=k_{y}k_{z}$
\\ 
 &$\psi^{(5+)}_{2}(\bm{k})=k_{z}k_{x}$
\\ 
 &$\psi^{(5+)}_{3}(\bm{k})=k_{x}k_{y}$
\\ 
\multicolumn{2}{c}{parity odd (triplet)}
\\ 
 $\Gamma^{-}_{1}$ &$\bm{d}^{(1-)}(\bm{k})=\hat{\bm{x}}k_{x}+\hat{\bm{y}}k_{y}+\hat{\bm{z}}k_{z}$
\\ 
$\Gamma^{-}_{2}$ &$\bm{d}^{(2-)}(\bm{k})=\hat{\bm{x}}k_{x}(k_{x}^{2}-k_{y}^{2})+\hat{\bm{y}}k_{y}+(k_{y}^{2}-k_{z}^{2})+\hat{\bm{z}}k_{z}(k_{z}^{2}-k_{x}^{2})$
\\ 
$\Gamma^{-}_{3}$ &$\bm{d}^{(3-)}_{1}(\bm{k})=2\hat{\bm{z}}k_{z}-\hat{\bm{x}}k_{x}-\hat{\bm{y}}k_{y}$
\\ 
 &$\bm{d}^{(3-)}_{2}(\bm{k})= \sqrt{3}(\hat{\bm{x}}k_{x}-\hat{\bm{y}}k_{y})$
\\ 
$\Gamma^{-}_{4}$ &$\bm{d}^{(4-)}_{1}(\bm{k})=\hat{\bm{y}}k_{z}-\hat{\bm{z}}k_{y}$
\\ 
 &$\bm{d}^{(4-)}_{2}(\bm{k})=\hat{\bm{z}}k_{x}-\hat{\bm{x}}k_{z}$
\\ 
 &$\bm{d}^{(4-)}_{3}(\bm{k})=\hat{\bm{x}}k_{y}-\hat{\bm{y}}k_{x}$
\\ 
$\Gamma^{-}_{5}$ &$\bm{d}^{(5-)}_{1}(\bm{k})=\hat{\bm{y}}k_{z}+\hat{\bm{z}}k_{y}$
\\ 
 &$\bm{d}^{(5-)}_{2}(\bm{k})=\hat{\bm{z}}k_{x}+\hat{\bm{x}}k_{z}$
\\ 
 &$\bm{d}^{(5-)}_{3}(\bm{k})=\hat{\bm{x}}k_{y}+\hat{\bm{y}}k_{x}$
\\ \hline
\multicolumn{2}{c}{$D_{6h}$}
\\  
\multicolumn{2}{c}{parity even (singlet)}
\\ 
 $\Gamma^{+}_{1}$ &$\psi^{(1+)}(\bm{k})=1$
\\ 
$\Gamma^{+}_{2}$ &$\psi^{(2+)}(\bm{k})=k_{x}k_{y}(k_{x}^{2}-3k_{y}^{2})(k_{y}^{2}-3k_{x}^{2})$
\\ 
$\Gamma^{+}_{3}$ &$\psi^{(3+)}(\bm{k})=k_{z}k_{x}(k_{x}^{2}-3k_{y}^{2})$
\\ 
$\Gamma^{+}_{4}$ &$\psi^{(4+)}(\bm{k})=k_{z}k_{y}(k_{y}^{2}-k_{x}^{2})$
\\ 
$\Gamma^{+}_{5}$ &$\psi^{(5+)}_{1}(\bm{k})=k_{x}k_{z}$
\\ 
 &$\psi^{(5+)}_{2}(\bm{k})=k_{y}k_{z}$
\\ 
$\Gamma^{+}_{6}$ &$\psi^{(6+)}_{1}(\bm{k})=k_{x}^{2}-k_{y}^{2}$
\\ 
 &$\psi^{(6+)}_{2}(\bm{k})=2k_{x}k_{y}$
\\ 
\multicolumn{2}{c}{parity odd (triplet)}
\\ 
 $\Gamma^{-}_{1}$ &$\bm{d}^{(1-)}(\bm{k})=\hat{\bm{x}}k_{x}+\hat{\bm{y}}k_{y},\hat{\bm{z}}k_{z}$
\\ 
$\Gamma^{-}_{2}$ &$\bm{d}^{(2-)}(\bm{k})=\hat{\bm{x}}k_{y}-\hat{\bm{y}}k_{x}$
\\ 
$\Gamma^{-}_{3}$ &$\bm{d}^{(3-)}(\bm{k})=\hat{\bm{z}}k_{x}(k_{x}^{2}-3k_{y}^{2}),k_{z}{(k_{x}^{2}-k_{y}^{2})\hat{\bm{x}}-2k_{x}k_{y}\hat{\bm{y}}}$
\\ 
$\Gamma^{-}_{4}$ &$\bm{d}^{(4-)}(\bm{k})=\hat{\bm{z}}k_{y}(k_{y}^{2}-3k_{x}^{2}),k_{z}{(k_{y}^{2}-k_{x}^{2})\hat{\bm{y}}-2k_{x}k_{y}\hat{\bm{x}}}$
\\ 
$\Gamma^{-}_{5}$ &$\bm{d}^{(5-)}_{1}(\bm{k})=\hat{\bm{x}}k_{z},\hat{\bm{z}}k_{x}$
\\ 
 &$\bm{d}^{(5-)}_{2}(\bm{k})=\hat{\bm{y}}k_{z},\hat{\bm{z}}k_{y}$
\\ 
$\Gamma^{-}_{6}$ &$\bm{d}^{(6-)}_{1}(\bm{k})=\hat{\bm{x}}k_{x}-\hat{\bm{y}}k_{y}$
\\ 
 &$\bm{d}^{(6-)}_{2}(\bm{k})=\hat{\bm{x}}k_{y}+\hat{\bm{y}}k_{x}$
\\ 
\hline
\multicolumn{2}{c}{$D_{4h}$}
\\ 
\multicolumn{2}{c}{parity even (singlet)}
\\ 
 $\Gamma^{+}_{1}$ &$\psi^{(1+)}(\bm{k})=1$
\\ 
$\Gamma^{+}_{2}$ &$\psi^{(2+)}(\bm{k})=k_{x}k_{y}(k_{x}^{2}-k_{y}^{2})$
\\ 
$\Gamma^{+}_{3}$ &$\psi^{(3+)}(\bm{k})=k_{x}^{2}-k_{y}^{2}$
\\ 
$\Gamma^{+}_{4}$ &$\psi^{(4+)}(\bm{k})=k_{x}k_{y}$
\\ 
$\Gamma^{+}_{5}$ &$\psi^{(5+)}_{1}(\bm{k})=k_{x}k_{z}$
\\ 
 &$\psi^{(5+)}_{2}(\bm{k})=k_{y}k_{z}$
\\ 
\multicolumn{2}{c}{parity odd (triplet)}
\\ 
 $\Gamma^{-}_{1}$ &$\bm{d}^{(1-)}(\bm{k})=\hat{\bm{x}}k_{x}+\hat{\bm{y}}k_{y},\hat{\bm{z}}k_{z}$
\\ 
$\Gamma^{-}_{2}$ &$\bm{d}^{(2-)}(\bm{k})=\hat{\bm{x}}k_{y}-\hat{\bm{y}}k_{x}$
\\ 
$\Gamma^{-}_{3}$ &$\bm{d}^{(3-)}(\bm{k})=\hat{\bm{x}}k_{x}-\hat{\bm{y}}k_{y}$
\\ 
$\Gamma^{-}_{4}$ &$\bm{d}^{(4-)}(\bm{k})=\hat{\bm{x}}k_{y}+\hat{\bm{y}}k_{x}$
\\ 
$\Gamma^{-}_{5}$ &$\bm{d}^{(5-)}_{1}(\bm{k})=\hat{\bm{x}}k_{z},\hat{\bm{z}}k_{x}$
\\ 
 &$\bm{d}^{(5-)}_{2}(\bm{k})=\hat{\bm{y}}k_{z},\hat{\bm{z}}k_{y}$
\\ 
\end{longtable}
Order parameters for $O_{h}$, $D_{6h}$, and $D_{4h}$ obtained from
GL and the change of their nodal structure following the introduction of
nonmagnetic
impurities are given in the following tables. For example,
$(1,\omega,\omega^{2})$ is a combination of expansion coefficients for
basis functions defined here, where
$\omega=e^{i2\pi/3}$. (a) means numbers of order parameters
being solutions of GL simultaneously. (b) means the
largest subgroup $G^{\prime}$ where the excitation energy is invariant ($\Gamma^{\prime +}_{1}$).
(c) means zeros in
$\bm{k}$-space immediately under the transition point $T_{c}$ in the
absence of impurities. $[k_{x},k_{y},k_{z}]$ in the table means a straight line generated by a vector
$(k_{x},k_{y},k_{z})$. $(k_{x},k_{y},k_{z})$ in the table means a plane which is perpendicular to
a vector $(k_{x},k_{y},k_{z})$ and goes through the origin of
$\bm{k}$-space. (d) means zeros below $T_{c}$ in the absence and the
presence of impurities. 
Nodes are generated where zeros and Fermi surface
intersect.
``none'' means nonexistence of node, and left arrow 
means that the nodes on the left-hand side
does not disappear. In the right edge, which representation appears is
written in the case of complex coefficients.
\\
\newpage
\renewcommand{\arraystretch}{1.5}
{\setlength{\tabcolsep}{10pt}
\renewcommand{\arraystretch}{1.5}
 \begin{longtable*}{cccc|c|c}
\caption{$O_{h}$}
\\
\hline \hline 
\multicolumn{1}{c}{$(O_{h})\Gamma$ }&
\multicolumn{1}{c}{$\psi(\bm{k}) / \bm{d}(\bm{k})$}&
\multicolumn{1}{c}{(a)}&
\multicolumn{1}{c}{(b)}&
\multicolumn{1}{c}{(c)}&
\multicolumn{1}{c}{(d)}\\ \hline \endhead \hline \endfoot
 $\Gamma^{+}_{1}$ & 1  &1
  &$O_{h}(\Gamma^{+}_{1})$ &none &none 
\\ \hline
$\Gamma^{+}_{2}$ &
  1  &1
  &$O_{h}(\Gamma^{+}_{2})$ &$(1,\pm 1,0)$,$(0,1,\pm 1)$,$(\pm 1,0,1)$& $\leftarrow$ 

\\ \hline
$\Gamma^{+}_{3}(1)$ & $(0,1)$  &3
  &$D_{4h}^{(0,0,1)}(\Gamma^{+}_{3})$ &$(1,\pm1,0)$&$\leftarrow$    
\\ \hline
$\Gamma^{+}_{3}(2)$ & $(1,0)$  &3
  &$D_{4h}^{(0,0,1)}(\Gamma^{+}_{1})$ &[1,1,1],[-1,1,1],[1,-1,1],[1,1,-1]&
  none
\\ \hline
$\Gamma^{+}_{3}(3)$ & $(1,i)$
  &2 &$O_{h}(\Gamma^{+}_{3})$ &$[1,1,1],[-1,1,1],[1,-1,1],[1,1,-1]$&$\leftarrow$  
\\ \hline
$\Gamma^{+}_{4}(1)$ & $(1,\omega,\omega^{2})$  &8 &$D_{3d}^{(1,1,1)}(\Gamma^{+}_{3})$
  &[1,1,1]&$\leftarrow$ 
\\  \hline
$\Gamma^{+}_{4}(2)$ & $(1,1,1)$  &4 &$D_{3d}^{(1,1,1)}(\Gamma^{+}_{2})$
  &$(1,-1,0)$,$(0,1,-1)
$,$(-1,0,1)$&$\leftarrow$ \\
 & & &
&[1,-1,0],[0,1,-1],[-1,0,1]&$\leftarrow$  \\
\hline
$\Gamma^{+}_{4}(3)$ & $(1,0,0)$  &3 &$D_{4h}^{(1,0,0)}(\Gamma^{+}_{2})$
  &$(0,1,\pm 1)$,$(0,0,1)
$,$(0,1,0)$&$\leftarrow$   \\ \hline
$\Gamma^{+}_{4}(4)$ & $(1,i,0)$  &6 &$D_{4h}^{(0,0,1)}(\Gamma^{+}_{5})$
  &$(0,0,1)$,$[0,0,1]
$&$\leftarrow$   \\
 & & &
& &  \\
 \hline
$\Gamma^{+}_{5}(1)$ & $(1,\omega,\omega^{2})$  &8 &$D_{3d}^{(1,1,1)}(\Gamma^{+}_{3})$
  &[1,1,1]
&$\leftarrow$   \\
& & &
&[1,0,0],[0,1,0],[0,0,1] &  \\ 
& & &
& &  \\ 
 \hline 
$\Gamma^{+}_{5}(2)$ & $(1,1,1)$  &4 &$D_{3d}^{(1,1,1)}(\Gamma^{+}_{1})$
  &[1,0,0],[0,1,0],[0,0,1]
&none   \\
& & &
& &  \\ 
\hline
$\Gamma^{+}_{5}(3)$ & $(1,0,0)$  &3 &$D_{4h}^{(0,0,1)}(\Gamma^{+}_{4})$
  &(0,1,0),(0,0,1)
&$\leftarrow$   \\ \hline
$\Gamma^{+}_{5}(4)$ & $(1,i,0)$  &6 &$D_{4h}^{(0,0,1)}(\Gamma^{+}_{5})$
  &$(0,0,1)$,$[0,0,1]$&$\leftarrow$ 
 \\
& & &
& &  \\ 
 \hline
$\Gamma^{-}_{1}$ & $1$  &1 &$O_{h}(\Gamma^{-}_{1})$
  &none&none 
 \\ \hline
$\Gamma^{-}_{2}$ &
  1  &1
  &$O_{h}(\Gamma^{-}_{2})$ &[1,0,0],[0,1,0],[0,0,1]& $\leftarrow$ 
 \\
  \hline
$\Gamma^{-}_{3}(1)$ & $(0,1)$  &3
  &$D_{4h}^{(0,0,1)}(\Gamma^{-}_{3})$ &[0,0,1]&$\leftarrow$    
\\ \hline
$\Gamma^{-}_{3}(2)$ & $(1,0)$  &3
  &$D_{4h}^{(0,0,1)}(\Gamma^{-}_{1})$ &none& none
\\
& & &
& &  \\
 \hline
$\Gamma^{-}_{3}(3)$ & $(1,i)$
  &2 &$O_{h}(\Gamma^{-}_{3})$ &none&none  
\\ \hline
$\Gamma^{-}_{4}(1)$ & $(1,\omega,\omega^{2})$  &8 &$D_{3d}^{(1,1,1)}(\Gamma^{-}_{3})$
  &none&none  
\\
& & &
& &  \\
& & &
& &  \\
 \hline
$\Gamma^{-}_{4}(2)$ & $(1,1,1)$  &4 &$D_{3d}^{(1,1,1)}(\Gamma^{-}_{2})$
  &none&none  \\
& & &
& &  \\
\hline
$\Gamma^{-}_{4}(3)$ & $(1,0,0)$  &3 &$D_{4h}^{(1,0,0)}(\Gamma^{-}_{2})$
  &none
&none  \\ \hline
$\Gamma^{-}_{4}(4)$ & $(0,-i,1)$  &6 &$D_{4h}^{(1,0,0)}(\Gamma^{-}_{5})$
  &none
&none  \\
& & &
& &  \\
\hline
$\Gamma^{-}_{5}(1)$ & $(1,\omega^{2},\omega)$  &8 &$D_{3d}^{(1,1,1)}(\Gamma^{-}_{3})$
 & none
&none \\
& & &
& &  \\
& & &
& &  \\ 
\hline
$\Gamma^{-}_{5}(2)$ & $(1,1,1)$  &4 &$D_{3d}^{(1,1,1)}(\Gamma^{-}_{1})$
  &none
&none  \\
& & &
& &  \\
 \hline
$\Gamma^{-}_{5}(3)$ & $(1,0,0)$  &3 &$D_{4h}^{(1,0,0)}(\Gamma^{-}_{4})$
 & [1,0,0]
&$\leftarrow$   \\ \hline
$\Gamma^{-}_{5}(4)$ & $(0,i,1)$  &6 &$D_{4h}^{(1,0,0)}(\Gamma^{-}_{5})$
  &none&none
 \\ 
& & &
& &  \\
\hline
 \end{longtable*}
}

{\setlength{\tabcolsep}{10pt}
 \begin{longtable*}{cccc|c|c}
\caption{$D_{6h}$}\\
\hline \hline
\multicolumn{1}{c}{$(D_{6h})\Gamma$}&
\multicolumn{1}{c}{$\psi(\bm{k}) / \bm{d}(\bm{k})$}&
\multicolumn{1}{c}{(a)}&
\multicolumn{1}{c}{(b)}&
\multicolumn{1}{c}{(c)}&
\multicolumn{1}{c}{(d)}\\ \hline \endhead \hline \endfoot
 $\Gamma^{+}_{1}$ & 1  &1
  &$D_{6h}(\Gamma^{+}_{1})$ &none &none 
\\ \hline
$\Gamma^{+}_{2}$ &
  1  &1
  &$D_{6h}(\Gamma^{+}_{2})$ &$(1,0,0)$,$(0,1,0)$,$(1,\pm \sqrt{3},0)$,$(\pm \sqrt{3},1,0)$& $\leftarrow$ 
\\ \hline
$\Gamma^{+}_{3}$ & 1 &1
  &$D_{6h}(\Gamma^{+}_{3})$ &$(1,0,0)$,(0,0,1),$(1,\pm \sqrt{3},0)$&$\leftarrow$  
\\ \hline
$\Gamma^{+}_{4}$ & 1&1 &$D_{6h}(\Gamma^{+}_{4})$
  &(0,1,0),(0,0,1),$(\pm \sqrt{3},1,0)$&$\leftarrow$  
\\ \hline
$\Gamma^{+}_{5}(1)$ & $(1,0)$  &3 &$D_{2h}(\Gamma^{+}_{3})$
  &(1,0,0),(0,0,1)
&$\leftarrow$   \\ 
& & &
& &  \\
\hline
$\Gamma^{+}_{5}(2)$ & $(0,1)$  &3 &$D_{2h}(\Gamma^{+}_{4})$
   &(0,1,0),(0,0,1)
&$\leftarrow$  \\
& & &
& &  \\
 \hline
$\Gamma^{+}_{5}(3)$ & $(1,i)$  &2 &$D_{6h}(\Gamma^{+}_{5})$
  &(0,0,1),[0,0,1]
&$\leftarrow$  \\ 
\hline
$\Gamma^{+}_{6}(1)$ & $(1,0)$  &3 &$D_{2h}(\Gamma^{+}_{1})$
  &$[0,0,1]$&none 
 \\
& & &
& &  \\
 \hline
$\Gamma^{+}_{6}(2)$ & $(0,1)$  &3 &$D_{2h}(\Gamma^{+}_{2})$
  &$(1,0,0)$,(0,1,0)&$\leftarrow$  
 \\ 
& & &
& &  \\
\hline
$\Gamma^{+}_{6}(3)$ & $(1,i)$  &2 &$D_{6h}(\Gamma^{+}_{6})$
  &[0,0,1]&$\leftarrow$ 
 \\ \hline
$\Gamma^{-}_{1}$ & 1  &1
  &$D_{6h}(\Gamma^{-}_{1})$ &none &$\leftarrow$ 
\\ \hline
$\Gamma^{-}_{2}$ &
  1  &1
  &$D_{6h}(\Gamma^{-}_{2})$ &none& $\leftarrow$ 
\\ \hline
$\Gamma^{-}_{3}$ & 1 &1
  &$D_{6h}(\Gamma^{-}_{3})$ &[0,0,1]&$\leftarrow$    
\\ \hline
$\Gamma^{-}_{4}$ & 1&1 &$D_{6h}(\Gamma^{-}_{4})$
  &[0,0,1]&$\leftarrow$ 
\\ \hline
$\Gamma^{-}_{5}(1)$ & $(1,0)$  &3 &$D_{2h}(\Gamma^{-}_{3})$
  &none
&none  \\
& & &
& &  \\
 \hline
$\Gamma^{-}_{5}(2)$ & $(0,1)$  &3 &$D_{2h}(\Gamma^{-}_{4})$
   &none
&none  \\ 
& & &
& &  \\
\hline
$\Gamma^{-}_{5}(3)$ & $(1,i)$  &2 &$D_{6h}(\Gamma^{-}_{5})$
  &none
&none  \\ \hline
$\Gamma^{-}_{6}(1)$ & $(1,0)$  &3 &$D_{2h}(\Gamma^{-}_{1})$
  &[0,0,1]&none 
 \\
& & &
& &  \\
 \hline
$\Gamma^{-}_{6}(2)$ & $(0,1)$  &3 &$D_{2h}(\Gamma^{-}_{2})$
  &[0,0,1]&none  
 \\
& & &
& &  \\
 \hline
$\Gamma^{-}_{6}(3)$ & $(1,i)$  &2 &$D_{6h}(\Gamma^{-}_{6})$
  &[0,0,1]&$\leftarrow$   
 \\ \hline
 \end{longtable*}
}

{\setlength{\tabcolsep}{10pt}
 \begin{longtable*}{cccc|c|c}
\caption{$D_{4h}$}\\
\hline \hline
\multicolumn{1}{c}{$(D_{4h})\Gamma$}&
\multicolumn{1}{c}{$\psi(\bm{k}) / \bm{d}(\bm{k})$}&
\multicolumn{1}{c}{(a)}&
\multicolumn{1}{c}{(b)}&
\multicolumn{1}{c}{(c)}&
\multicolumn{1}{c}{(d)}\\ \hline \endhead \hline \endfoot
 $\Gamma^{+}_{1}$ & 1  &1
  &$D_{4h}(\Gamma^{+}_{1})$ &none &none 
\\ \hline
$\Gamma^{+}_{2}$ &
  1  &1
  &$D_{4h}(\Gamma^{+}_{2})$ &\phantom{0000}$(1,0,0)$,$(0,1,0)$,$(1,\pm 1,0)$\phantom{0000}& $\leftarrow$ 
\\ \hline
$\Gamma^{+}_{3}$ & 1 &1
  &$D_{4h}(\Gamma^{+}_{3})$ &$(1,\pm 1,0)$&$\leftarrow$  
\\ \hline
$\Gamma^{+}_{4}$ & 1&1 &$D_{4h}(\Gamma^{+}_{4})$
  &(1,0,0),(0,1,0)&$\leftarrow$ 
\\ \hline
$\Gamma^{+}_{5}(1)$ & $(1,i)$  &2 &$D_{4h}(\Gamma^{+}_{5})$
  &[0,0,1],(0,0,1)
&$\leftarrow$  \\ \hline
$\Gamma^{+}_{5}(2)$ & $(1,1)$  &2 &$D_{2h}^{(\rm{diagonal})}(\Gamma^{+}_{3})$
   &(0,0,1),(1,1,0)
&$\leftarrow$   \\ 
\hline
$\Gamma^{+}_{5}(3)$ & $(1,0)$  &2 &$D_{2h}(\Gamma^{+}_{3})$
  &(1,0,0),(0,0,1)
&$\leftarrow$  \\ \hline
 $\Gamma^{-}_{1}$ & 1  &1
  &$D_{4h}(\Gamma^{-}_{1})$ &none &none 
\\ \hline
$\Gamma^{-}_{2}$ &
  1  &1
  &$D_{4h}(\Gamma^{-}_{2})$ &none& none
\\ \hline
$\Gamma^{-}_{3}$ & 1 &1
  &$D_{4h}(\Gamma^{-}_{3})$ &[0,0,1]&$\leftarrow$   
\\ \hline
$\Gamma^{-}_{4}$ & 1&1 &$D_{4h}(\Gamma^{-}_{4})$
  &[0,0,1]&$\leftarrow$ 
\\ \hline
$\Gamma^{-}_{5}(1)$ & $(1,i)$  &2 &$D_{4h}(\Gamma^{-}_{5})$
  &none
&none  \\ \hline
$\Gamma^{-}_{5}(2)$ & $(1,1)$  &2 &$D_{2h}^{(\rm{diagonal})}(\Gamma^{-}_{3})$
   &none
&none  \\ \hline
$\Gamma^{-}_{5}(3)$ & $(1,0)$  &2 &$D_{2h}(\Gamma^{-}_{3})$
  &none
&none   \\ \hline
 \end{longtable*}
}

Here, $(0,0,1)$ of $D_{4h}^{(0,0,1)}(\Gamma^{3+})$ means as
follows. $O_{h}$ has three kinds of subgroup $D_{4h}$ which have
different four times rotation axis each other, for example,
$D_{4h}^{(0,0,1)}(\Gamma^{+}_{3})$ is a $D_{4h}$ which has $(0,0,1)$ as
four times rotation axis. Although such subgroups are different for every
order parameter, we write only subgroups for representative order
parameters. 

The largest subgroup where the quasiparticle energy of a given order
parameter is invariant can be obtained as follows\cite{koster}. First, examine all
maximal subgroups whether the excitation energy is invariant or not under
the subgroups. If there exits, that is the largest subgroup. If not,
examine all second maximal subgroups, and so on. The maximal subgroup
obtained like this is also the largest subgroup as mentioned in section
IV. 
\\

\section{Expansion of Order Parameters of Complex Coefficients with respect to $G^{\prime}$ where the Quasiparticle Energy of Clean Superconductor is Invariant}
We examined $\{c_{i}^{\prime}\}$ of
$\sum_{i}c_{i}^{\prime}\psi_{i}^{\prime(\beta)}=\sum_{i}c_{i}\psi_{i}^{(\alpha)}$.
$\psi_{i}^{\prime(\beta)}$ means that it belongs
to $i$th basis of $\beta$th irreducible representation of $G^{\prime}$
where the quasiparticle energy of clean superconductor is invariant.
These results can be understood from the GL theory.

 \begin{tabular}{|c|}\hline
$O_{h}$
\\ \hline 
 \end{tabular}

(2D)even

$\Gamma^{+}_{3}(3)$(subgroup $G^{\prime}=O_{h}$)
\begin{align}
&k_{x}^{2}+\omega
k_{y}^{2}+\omega^{2}k_{z}^{2}\nonumber \\
=&\frac{1}{2}(2k_{z}^{2}-k_{x}^{2}-k_{y}^{2})+\frac{i}{2}(\sqrt{3}(k_{x}^{2}-k_{y}^{2}))\nonumber \\
 =&\psi_{1}^{\prime(3+)}+i\psi_{2}^{\prime(3+)}
\end{align}
\\

(2D)odd

$\Gamma^{-}_{3}(3)$(subgroup $G^{\prime}=O_{h}$)
\begin{align}
\hat{\bm{x}}k_{x}+\omega
\hat{\bm{y}}k_{y}+\omega^{2}\hat{\bm{z}}k_{z}=\psi_{1}^{\prime(3-)}+i\psi_{2}^{\prime(3-)}
\end{align}
\\

(3D)even

$\Gamma^{+}_{4}(1)$(subgroup $G^{\prime}=D_{3d}$ which has (1,1,1) of $O_{h}$
as a three times rotation axis)
\begin{align}
&k_{y}k_{z}(k_{y}^{2}-k_{z}^{2})+\omega
k_{z}k_{x}(k_{z}^{2}-k_{x}^{2})+\omega^{2}
k_{x}k_{y}(k_{x}^{2}-k_{y}^{2})\notag\\
=&\frac{3-i\sqrt{3}}{4}(k_{x}+k_{y})k_{z}(k_{x}^{2}-k_{x}k_{y}+k_{y}^{2}-k_{z}^{2})+\notag\\
&\frac{-1-i\sqrt{3}}{4}(k_{x}^{3}(2k_{y}+k_{z})-k_{y}k_{z}(k_{y}^{2}-k_{z}^{2})-k_{x}(2k_{y}^{3}+k_{z}^{3}))\notag\\
=&\psi_{1}^{\prime(3+)}+
i\psi_{2}^{\prime(3+)}\notag\\
\end{align}
\\

$\Gamma^{+}_{4}(4)$(subgroup $G^{\prime}=D_{4h}$ which has (0,0,1) of
$O_{h}$ as a four times rotation axis)
\begin{align}
&k_{y}k_{z}(k_{y}^{2}-k_{z}^{2})+i k_{z}k_{x}(k_{z}^{2}-k_{x}^{2})\notag\\
=&\psi_{1}^{\prime(5+)}+i\psi_{2}^{\prime(5+)}
\end{align}
\\

$\Gamma^{+}_{5}(1)$(subgroup $G^{\prime}=D_{3d}$ which has (1,1,1) of
$O_{h}$ as a three times rotation axis)
\begin{align}
&k_{y}k_{z}+\omega
k_{z}k_{x}+\omega^{2}
k_{x}k_{y}\notag\\
=&-\frac{1+i\sqrt{3}}{4}(2k_{x}k_{y}-k_{x}k_{z}-k_{y}k_{z})\nonumber \\
&-i\frac{\sqrt{3}-i\sqrt{3}}{4}(\sqrt{3}(k_{x}-k_{y})k_{z})\nonumber\\
=&\psi_{1}^{\prime(3+)}+i
 \psi_{2}^{\prime(3+)}\nonumber\\
\end{align}
\\

$\Gamma^{+}_{5}(4)$(subgroup $G^{\prime}=D_{4h}$ which has (0,0,1) of
$O_{h}$ as a four times rotation axis)
\begin{align}
&k_{y}k_{z}+i k_{z}k_{x}\notag\\
 =&\psi_{1}^{\prime(5+)}+i
 \psi_{2}^{\prime(5+)}
\end{align}
\\

(3D)odd

$\Gamma^{-}_{4}(1)$(subgroup $G^{\prime}=D_{3d}$ which has (1,1,1) of
$O_{h}$ as a three times rotation axis)
\begin{align}
&\hat{\bm{x}}(\omega k_{y}-\omega^{2} k_{z})+\hat{\bm{y}}(k_{z}-\omega
 k_{x})+\hat{\bm{z}}(\omega^{2} k_{x}-k_{y})
\notag\\
=&\frac{3+i\sqrt{3}}{4}(\hat{\bm{x}}(-k_{z})+\hat{\bm{y}}(-k_{z})+\hat{\bm{z}}(k_{x}+k_{y}))+\notag\\
&\frac{-1+i\sqrt{3}}{4}(\hat{\bm{x}}(-2k_{y}-k_{z})+\hat{\bm{y}}(2k_{x}+k_{z})+\hat{\bm{z}}(k_{x}-k_{y}))\notag\\
=&\psi_{1}^{\prime(3-)}+i\psi_{2}^{\prime(3-)}\notag\\
\end{align}
\\

$\Gamma^{-}_{4}(4)$(subgroup $G^{\prime}=D_{4h}$ which has (1,0,0) of
$O_{h}$ as a four times rotation axis)
\begin{align}
&\hat{\bm{x}}(k_{y}+i k_{z})+\hat{\bm{y}}(-k_{x})+\hat{\bm{z}}(-i
 k_{x})
\notag\\
=&\psi_{1}^{\prime(5-)}+i\psi_{2}^{\prime(5-)}
\end{align}
\\

$\Gamma^{-}_{5}(1)$(subgroup $G^{\prime}=D_{3d}$ which has (1,1,1) of
$O_{h}$ as a three times rotation axis)
\begin{align}
&\hat{\bm{x}}(\omega k_{y}+\omega^{2} k_{z})+\hat{\bm{y}}(k_{z}+\omega
 k_{x})+\hat{\bm{z}}(\omega^{2} k_{x}+k_{y})
\notag\\
=&\frac{-1+i\sqrt{3}}{4}(\hat{\bm{x}}(k_{z}-2k_{y})+\hat{\bm{y}}(k_{z}-2k_{x})+\hat{\bm{z}}(k_{x}+k_{y}))+\notag\\
&\frac{\sqrt{3}+i}{4}(\sqrt{3}(\hat{\bm{x}}k_{z}+\hat{\bm{y}}(-k_{z})+\hat{\bm{z}}(k_{x}-k_{y})))\notag\\
=&\psi_{1}^{\prime(3-)}-i\psi_{2}^{\prime(3-)}\notag\\
\end{align}
\\

$\Gamma^{-}_{5}(4)$(subgroup $G^{\prime}=D_{4h}$ which has (0,0,1) of
$O_{h}$ as a four times rotation axis)
\begin{align}
&\hat{\bm{x}}(k_{y}+i k_{z})+\hat{\bm{y}}k_{x}+\hat{\bm{z}}(i k_{x})
\notag\\
=&\psi_{1}^{\prime(5-)}+i\psi_{2}^{\prime(5-)}
\end{align}
\\

 \begin{tabular}{|c|}\hline
$D_{6h}$
\\ \hline 
 \end{tabular}

(2D)even

$\Gamma^{+}_{5}(3)$(subgroup $G^{\prime}=D_{6h}$)
\begin{align}
&k_{z}(k_{x}+ik_{y})=\psi_{1}^{(5+)}+i\psi_{2}^{(5+)}
\end{align}
\\

$\Gamma^{+}_{6}(3)$(subgroup $G^{\prime}=D_{6h}$)
\begin{align}
&(k_{x}+ik_{y})^{2}=\psi_{1}^{(6+)}+i\psi_{2}^{(6+)}
\end{align}
\\

(2D)odd

$\Gamma^{-}_{5}(3)$(subgroup $G^{\prime}=D_{6h}$)
\begin{align}
&\hat{\bm{x}}k_{z}+i\hat{\bm{y}}k_{z}=\psi_{1}^{(5-)}+i\psi_{2}^{(5-)}
\end{align}
\\

$\Gamma^{-}_{6}(3)$(subgroup $G^{\prime}=D_{6h}$)
\begin{align}
&(\hat{\bm{x}}+i\hat{\bm{y}})(k_{x}+ik_{y})=\psi_{1}^{(6-)}+i\psi_{2}^{(6-)}
\end{align}
\\

 \begin{tabular}{|c|}\hline
$D_{4h}$
\\ \hline 
 \end{tabular}

(2D)even

$\Gamma^{+}_{5}(1)$(subgroup $G^{\prime}=D_{4h}$)
\begin{align}
&k_{z}(k_{x}+ik_{y})=\psi_{1}^{(5+)}+i\psi_{2}^{(5+)}
\end{align}
\\

(2D)odd

$\Gamma^{-}_{5}(1)$(subgroup $G^{\prime}=D_{4h}$)
\begin{align}
&\hat{\bm{x}}k_{z}+i\hat{\bm{y}}k_{z}=\psi_{1}^{(5-)}+i\psi_{2}^{(5-)}
\end{align}
\\

\end{document}